\documentclass[prd,nofootinbib,twocolumn]{revtex4}
\usepackage{graphics}
\usepackage{bm}
\usepackage{float}
\usepackage{subfigure}

\begin{document}

\title{Intercommutation of Semilocal Strings and Skyrmions}

\author{Pablo Laguna}
\affiliation{Department of Astronomy \& Astrophysics, IGPG, CGWP\\
Penn State University, University Park, PA 16802}

\author{Vishnu Natchu}
\affiliation{Center for Relativity, Department of Physics\\
The University of Texas at Austin, Austin, TX 78712}

\author{Richard A. Matzner}
\affiliation{Center for Relativity, Department of Physics\\
The University of Texas at Austin, Austin, TX 78712}

\author{Tanmay Vachaspati}
\affiliation{CERCA, Department of Physics, Case Western Reserve University,
10900 Euclid Avenue, Cleveland, OH 44106-7079}

\begin{abstract}
\noindent
We study the intercommuting of semilocal strings and Skyrmions,
for a wide range of internal parameters, velocities and intersection
angles by numerically evolving the equations of motion. We find that
the collisions of strings and strings, strings and Skyrmions, and
Skyrmions and Skyrmions, all lead to intercommuting for a wide range 
of parameters. Even the collisions of unstable Skyrmions and strings
leads to intercommuting, demonstrating that the phenomenon of
intercommuting is very robust, extending to dissimilar
field configurations that are not stationary solutions. Even more
remarkably,
at least for the semilocal $U(2)$ formulation considered here, all
intercommutations trigger a reversion to U(1) Nielsen-Olesen strings.
\end{abstract}

\maketitle

Physical systems can have a large number of spontaneously broken
symmetries, some of which may be local (gauged) while others may
be global. In cases where the symmetries are only partially gauged,
or gauged with unequal strengths, semilocal strings may be present
\cite{Vachaspati:1991dz,Hindmarsh:1991jq,Preskill:1992bf,Achucarro:1999it}.
If the gauge couplings are widely disparate, the strings may also 
be stable. Semilocal string solutions exist in the standard electroweak 
model, but the $SU(2)$ and $U(1)$ gauge couplings are close enough 
that the solutions are unstable \cite{Vachaspati:1992fi,James:1992zp}. 
Recently twisted and current-carrying semilocal string  solutions have 
been discovered that may be stable for other values of parameters 
\cite{Forgacs:2005sf,Forgacs:2006pm}.
Semilocal strings arise in supersymmetric QCD if the number of
flavors is larger than the order of the symmetry group \cite{Shi06}.
In string theory, they can arise as field theoretic realizations
of cosmic D-strings \cite{Blanco-Pillado:2005xx}, in which
case they may play a role in cosmology. The formation of semilocal
strings in a phase transition has been studied in
Refs.~\cite{Achucarro:1998ux,Ach06}, in a cosmological setting
in Refs.~\cite{Benson:1993at,Urrestilla:2004eh}, and in superstring
cosmology \cite{Dasgupta:2004dw}.



\begin{figure}
\subfigure[Initial configuration]{\scalebox{0.40}{\includegraphics{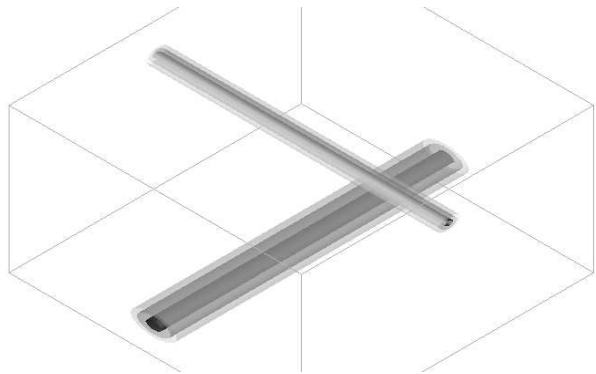}}}
\subfigure[Late configuration]{\scalebox{0.40}{\includegraphics{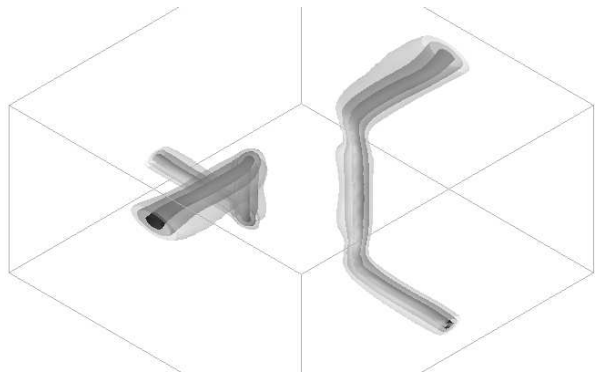}}}
\caption{ Field $\sigma$ for a collision of $q_0 = 1$ (``top") and
  $q_0 = 3$ semilocal strings, at $\Theta = 90^\circ$, with collision
  velocity $V= 0.9$. The contours are $1-| \sigma|^2 $ at values $0.1,
  0.2, 0.5$. Note that $|\sigma|^2 + | \phi | ^2 = 1 $ is the
  vacuum. The strings reconnect.}
\label{fig1}
\end{figure}

In the case of topological gauged
\cite{Mat88,Moriarty:1988qs,Laguna:1989hn} and global
\cite{Shellard:1988ki} $U(1)$ strings, numerical evolution of the
field theory equations show that colliding strings intercommute
(Fig.~\ref{fig1}) for almost \cite{Bettencourt:1996qe} any scattering
angle and velocity. Thus the probability of
intercommuting is {\it unity} for topological strings.
Qualitative arguments \cite{Shellard:1988ki,Vachaspati:1989xg}
have been given in an effort to understand intercommutation of
topological strings.
In the context of string theory cosmic strings and QCD strings,
intercommuting probabilities have been calculated in
Refs.~\cite{Jackson:2004zg,Hashimoto:2005hi} within various
approximations.

\begin{figure}
\subfigure[Initial configuration]{\scalebox{0.40}{\includegraphics{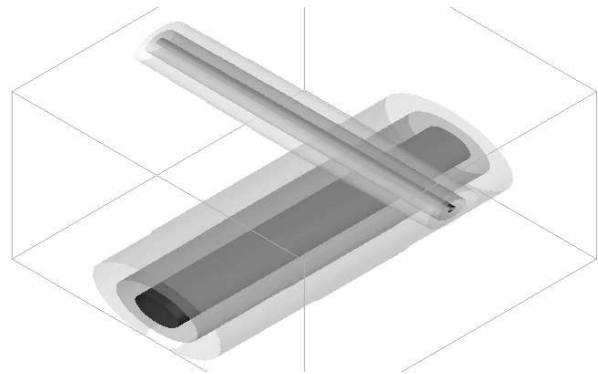}}}
\subfigure[Late configuration]{\scalebox{0.40}{\includegraphics{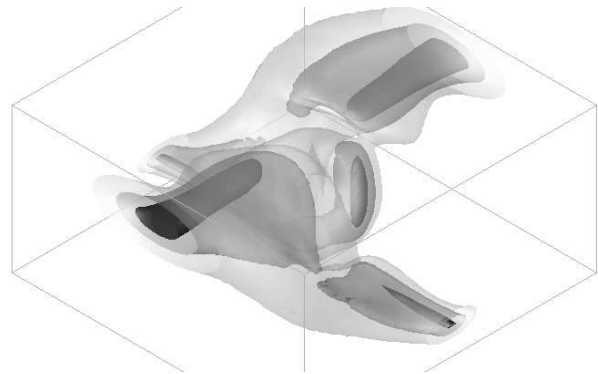}}}
\caption{Coincident configurations of the field $\phi$ in the collision of
FIG 1. The contours are $|\phi|^2$ at $0.2, 0.3, 0.5$. The value $\phi = 0$
is an NO string. It can be seen that the configurations reverts to NO form
after the collision.
This is in spite of the fact our boundary conditions hold
the boundary values of the strings in their original $q_0 = 1$ and $q_0 = 3$
state. The transition from the initial value of $\phi$ to the value of zero
means we have two NO strings near the center terminating on the (outward
moving) ends of the $q\ne 0$ semilocal strings.
}
\label{fig2}
\end{figure}

In this paper we study the interactions of {\it semilocal strings},
which are topological structures embedded in the solution
space of the partially gauged theory, and the related
{\it semilocal Skyrmions} which are fat and fuzzy objects (still
with one ``long", {\it i.e.} stringlike dimension) on which
strings can terminate, much like strings ending on
monopoles.  (We shall show below (Figs \ref{fig1}, \ref{fig2}) however,
that the termination is dynamic, moving along the string.)
The scalar field in a
Skyrmion can be arbitrarily close to its vacuum expectation
value everywhere, in contrast to a string in which the
scalar field magnitude vanishes (which is a nonvacuum value) on
some curve. We review below that the embedded topological
strings have precisely the dynamics of $U(1)$
({\it Nielsen-Olesen} (NO) \cite{Nielsen:1973cs}) strings.
Thus semilocal strings
will intercommute on collision with each other. However, the
interaction of {\em different} objects such as semilocal strings
with Skyrmions, or Skyrmions with Skyrmions, is a new consideration.
Remarkably we find that intercommuting is extremely robust --
strings intercommute with Skyrmions, and Skyrmions intercommute
with other Skyrmions. Indeed, even in the parameter regime where
the solutions are unstable, we see intercommuting. This establishes,
for the first time, that intercommuting is not particular to string
solutions, but can happen more generally between any two field
configurations. Even more remarkably,
at least for the semilocal $U(2)$ formulation considered here, all
intercommutations trigger a reversion to U(1) NO strings.



The semilocal model we shall study is the bosonic part of the
electroweak model with weak mixing angle $\theta_w =\pi/2$.
The Lagrangian is:
\begin{equation}
L = |(\partial_\mu - iY_\mu )  \Phi |^2  -
             {1 \over 4} Y_{\mu \nu} Y^{\mu \nu}  -
  \frac{\beta}{2} ( \Phi^\dag \Phi - 1 )^2
\label{semilocalaction}
\end{equation}
where $\Phi^T = ( \phi, \sigma )$ is an $SU(2)$ doublet,
$Y_{\mu} $ is a $U(1)$ gauge potential and
$Y_{\mu \nu}= \partial_{\mu}Y_{\nu} - \partial_{\nu}Y_{\mu}$
its field strength. The only parameter in the model is $\beta$.

The semilocal model has $[SU(2)\times U(1)_Y]/Z_2$ symmetry
and only the $U(1)_Y$ is gauged. If $\Phi^T$ gets a vacuum
expectation value along the direction $(0, 1)$
the symmetry is broken down to $U(1)_Q$ where
$Q = (T_3 + Y)/2 = {\rm diag}(1,0)$,
$T_3 = {\rm diag}(1,-1)/2$, and $Y = {\rm diag}(1,1)/2$.

The NO string is an embedded string
solution to Eq.(\ref{semilocalaction}):
\cite{Vachaspati:1992pi,Barriola:1993fy}
$\Phi^T = (0,f e^{i\theta})$ and
$Y_\theta = (v/r)\, {\hat e}_\theta$
in cylindrical coordinates $(r,\theta ,z)$ with $f(r)$
and $v(r)$ profile functions identical to those for the
NO string \cite{Nielsen:1973cs}.
Further, the dynamics and interactions of fields living in the
embedded subspace -- lower component
of $\Phi$ and $Y_\mu$ -- are given exactly by the Abelian-Higgs
model and so these behave in {\it exactly} the
same way that NO strings of the Abelian-Higgs model do,
and intercommute regardless of collisional
velocity and angle
\cite{Mat88,Moriarty:1988qs,Laguna:1989hn}. However, in
this subspace of the field theory, excitations that can cause
the string to terminate in Skyrmions are absent. So to study
the consequences of semilocality for string interactions, we need
to go beyond the embedded Abelian-Higgs model.

It has been shown that the NO string, as embedded in this model,
is stable for $\beta < 1$,
unstable for $\beta > 1$ and neutrally stable for $\beta =1$
\cite{Hindmarsh:1991jq}.  This means that if we start with a string-like
configuration that involves degrees of freedom outside the embedded
Abelian-Higgs model, it will relax into the string solution for
$\beta < 1$ and go further out of the embedding space for $\beta > 1$.
Hence the most suitable choice of parameter for studying the effects
of semilocality on the interactions of strings is precisely at
$\beta =1$ and this is the value that we adopt for most of
our analysis.

At $\beta =1$ there is a class of solutions that are degenerate
in energy and that can be written as \cite{Gibbons:1992gt}:
\begin{equation}
\Phi = \frac{1}{\sqrt{r^2 + |q_0|^2}} \pmatrix{q_0\cr r e^{i\theta}}
          \exp \biggl [ \frac{1}{2} u(r; |q_0|) \biggr ]
\end{equation}
where $u = \ln |\Phi |^2$ is the solution to:
\begin{equation}
\nabla^2 u + 2(1-e^u) = \nabla^2 \ln (r^2 + |q_0|^2 )
\end{equation}
and $u \rightarrow 0$ as $r \rightarrow \infty$. The complex parameter
$q_0$ is a coordinate on the space of solutions. At $q_0=0$, the
solution is the embedded NO string.
For other values of $|q_0|$ the magnetic field and the energy density
is more spread out and the solutions are called ``semilocal
Skyrmions'' \cite{Benson:1993at} or, generically, semilocal strings
with particular values of $q_0$. The phase of $q_0$ labels a
Goldstone mode -- for example, the phase can be pulled out as
an overall factor of the field $\Phi$, and be compensated for in the lower
component by shifting $\theta$. The relative phases of $q_0$ for
two strings (called the relative ``color'' in \cite{Achucarro:1992hs})
can still be significant but the phase difference is expected to be quickly
radiated away as the two
strings collide \cite{Achucarro:1992hs}, so it is sufficient to take
$q_0$ to be real. In fact, as described below, our present simulations
show that $q_0 \rightarrow 0$ in the aftermath of the collision.

We now set up two semilocal strings with (real) $q_0$ values
$q_0^{(1)}$ and $q_0^{(2)}$. The strings are then boosted so that they
approach each other at relative velocity, $V$ (in units of $c$), and angle
$\Theta$.
Numerical evolution of the equations then tells us whether the
strings intercommute or pass through. 

Initial data and boundary conditions are based on suitable superpositions of
single, boosted, static string solutions. Static solutions
$\Phi^T = (g, f e^{i\theta})$ and
$Y_\theta = (v/r) {\hat e}_\theta$ for
semilocal strings
with $\beta=1$ can be found from:
\begin{eqnarray}
f' + \frac{v-1}{r}\,f &=& 0 \label{eq:f}\\
v' + r\,\left[f^2\,\left(1+\frac{q_o^2}{r^2}+g^2-1\right)\right] &=&
0\label{eq:v2}\,.
\end{eqnarray}
where $f$, $g$ and $v$ are functions of $r$ only and $g = q_o\,f/r$.
To set up initial data and boundary conditions one could
numerically solve Eqs.~(\ref{eq:f}) and (\ref{eq:v2}) and construct a
table to be used for the superposition of string solutions.
We choose instead parametrized approximate solutions to the
equations. That is,
\begin{eqnarray}
f &=& \left[1 - a_1 \,\left(\frac{r}{1+r^2}\right)^2\right]\,f_{NO}
\end{eqnarray}
where
$f_{NO} = [\tanh{(r\,R_f/2)}]^{E_f}$,
$R_f = 1+(a_2-1)\,e^{-r/r_o}$ and
$E_f = a_3 + (1-a_4)\,e^{-r/r_o}$
with similar expressions for $v$. The parameters in these expressions are
found by minimizing the total energy
for a single string.

Note that the fields for semilocal strings with non-zero $q_0$ approach
their vacuum values only as power laws ($r^{-2}$ for $\Phi$, and $r^{-4}$
for the magnetic field strength). This is to be contrasted with the
exponential
approach for NO strings (which is the $q_0 = 0$ case). Nonetheless these
polynomial fall-offs
are still fast enough that we anticipate being able to construct initial
data by superposing fields for widely separated strings, with minimal change
in the total energy. The scheme is remarkably straightforward. (An early
reference to a similar scheme for NO strings is
Abrikosov \cite{Abrikosov}. See also Matzner \cite{Mat88}.)

%
Next we consider how to superpose the scalar fields of two
well-separated strings, labeled by $i=1,2$. Let the two string
scalar field configurations be
$\Phi_1^T = (\phi_1, \sigma_1)$
and
$\Phi_2^T = (\phi_2, \sigma_2),$
respectively.
Since the $\phi_i$ vanish as a power law
at infinity, and $\sigma_i$ go to unity, we superpose
the scalar field string configurations using the scheme
$\Phi^T = (\phi_{1} + \phi_{2},\sigma_{1}\sigma_{2})$.
Because the axial gauge field $Y_\mu$ satisfies an equation
which is linear,
the superposition of $Y_\mu$ is taken to be
$Y_\mu = Y_\mu^{(1)} + Y_\mu^{(2)}$.

The actual data are set as follows.
With the energy minimization procedure described above,
we start by constructing two single string solutions
with parameters $q_o^{(1)}$ and $q_o^{(2)}$ in coordinates $(x, y, z)$.
The strings are parallel to the $z$ axis and pass through the origin.
We henceforth assume that $Y_{\mu}$ is written in the Lorentz gauge:
$\partial_\mu Y^\mu = 0$. This gauge is linear and Lorentz invariant.
The string $q_o^{(2)}$ is then
rotated around the $x$-axis by an angle $\Theta$, with $\Theta = 0$
(parallel strings) and
$\Theta = \pi$ (antiparallel strings).
Next, both strings are off-set along the $x$-axis a distance
$x_o^{(1)}=-D/2$ and $x_o^{(2)}=D/2$, respectively.
Finally, the strings are Lorentz boosted toward each other with velocity
$V$, so that the string intersection will occur in the middle of the
computational cube. Initial time derivatives are computed to second order
in the timestep by repeating the data
construction process at time $\Delta t/2$ into the future and $\Delta
t/2$ into the past and taking $1/\Delta t$ times the
difference of the two field configurations.

The boundary data are naturally set as part of the initial setup process.
In subsequent time steps,
the boundary values are computed as if this procedure were repeated, with
the strings advanced at the boundary as
if in free motion with the initial velocity. This means that even after the
collision in the center
of the cube, the boundaries continue to act as if the strings had passed
through one another,
regardless whether they did in fact pass through one another, or
reconnected.
However our computational domains are
large enough that the region where the strings collide is
causally disconnected from the boundaries of the computational domain.

Our numerical code solves the field equations of motion for
$\Phi^T = ( \phi, \sigma )$ and
$Y_{\mu} $ as a coupled set of first-order differential equations in time:
\begin{eqnarray}
\partial_t \eta_A &=&
       \nabla^2 \phi_A - Y^2\,\phi_A  - \partial_{\phi_A}U\nonumber \\
       &&-2\,\epsilon_{AB}\,(Y^t\,\eta_B + Y^i\,\partial_i\phi_B) \\
\partial_t \kappa_A &=&
  \nabla^2 \sigma_A - Y^2\,\sigma_A  - \partial_{\sigma_A}U\nonumber \\
       &&-2\,\epsilon_{AB}\,(Y^t\,\kappa_B + Y^i\,\partial_i\sigma_B) \\
\partial_t W_\mu &=&
        \nabla^2 Y_\mu\nonumber - Y_\mu\,(\phi^2 + \sigma^2)\nonumber\\
       &&-\epsilon_{AB}\,(\phi_A\,\partial_\mu\phi_B +
\sigma_A\,\partial_\mu\sigma_B) \,,
\label{partialW}
\end{eqnarray}
where
$\partial_t \phi_A = \eta_A,$
$\partial_t \sigma_A = \kappa_A$ and
$\partial_t Y_\mu = W_\mu$.
The subscripts $A,B = 1,2$ denote respectively real and imaginary
parts,
$\epsilon_{AB}$ is the completely antisymmetric tensor
in two indices with $\epsilon_{12} = 1$ and $U= \beta(\phi^2 + \sigma^2
-1)^2/2$.
We use a second order accurate, both in space and time, discretization. The
temporal updating is done via a standard leap-frog method.

Our first set of numerical experiments were done
with strings of the same internal
parameter $q_0$
( $q_0^{(2)}=q_0^{(1)}$). We first verified that $q_0=0$ strings (NO
strings)
reconnect, as had been found previously. We also considered collisions
between
$q_0=1$ strings.
Just like NO strings, these $q_0=1$
semilocal strings always intercommuted for the set of velocities
$V \in \{0.1,0.5,0.9\} \equiv \{\bf V\}$ and all intersection angles tested:
\{$30^\circ,45^\circ,60^\circ,90^\circ,120^\circ,150^\circ$\}  $\equiv \{\bf
\Theta\}$.
Next, we consider the case
$q_0^{(2)} \ne q_0^{(1)}$.
Our simulations now depend on four parameters: $V$, $\Theta$, $q_0^{(1)}$
and $q_0^{(2)} - q_0^{(1)}$. To reduce the parameter space,
we only considered $q_0^{(1)}=1$, and $q_0^{(2)} \in
\{0.0,0.5,0.9, 1.0, 1.1, 2.0, 3.0\}$. In these cases also we find that
for all velocities
$V \in
\{\bf V\}$ and all angles $\Theta \in
\{\bf \Theta\}$,
reconnection
always occurs. The dynamics involved in the reconnection are interesting.
In particular, the value of  $q_0$ for both strings
appears to relax (by radiation)
toward $q_0=0$. The intercommutation converts these to NO strings!
Figures \ref{fig1} and \ref{fig2} show frames from simulations
of a right angle collision at $V= 0.9$, between
$q_0 = 1$ and $q_0 = 3$ semilocal strings
showing this behavior.  We also considered collisions between unstable
($\beta > 1$) Skyrmions. Again, intercommutation always occurred, and the
reversion to NO always occured. The full range of the collision
simulations is available online at
{\tt http://gravity.psu.edu/numrel/strings/}

We have numerically studied collisions of strings and Skyrmions
in the semilocal model. Our results demonstrate that collisions of any two
objects in this model, whether they are identical or not, whether they are
stable or unstable configurations, leads to intercommuting for any set
of collisional parameters. Additionally, the simulations all show a
triggered collapse of the interacting objects toward a NO configuration.
During the collision we also observe a large amount of energy
lost in radiation. This may provide an additional signature for
cosmological semilocal strings.

\begin{acknowledgments}
This work was supported in part by the U.S. Department of Energy and
NASA at Case to TV, by NSF grant PHY-0244788 to PL and the Center for
Gravitational Wave Physics funded by the NSF under Cooperative Agreement
PHY-0114375, and by NSF PHY 0354842 and NASA
NNG04Gl37G to Richard Matzner.
\end{acknowledgments}


\begin{thebibliography}{}

\bibitem{Vachaspati:1991dz}
 T.~Vachaspati and A.~Achucarro,
 Phys.\ Rev.\ D {\bf 44}, 3067 (1991).

\bibitem{Hindmarsh:1991jq}
 M.~Hindmarsh,
 Phys.\ Rev.\ Lett.\  {\bf 68}, 1263 (1992).

\bibitem{Preskill:1992bf}
 J.~Preskill,
 Phys.\ Rev.\ D {\bf 46}, 4218 (1992)
 [arXiv:hep-ph/9206216].

\bibitem{Achucarro:1999it}
 A.~Achucarro and T.~Vachaspati,
 Phys.\ Rept.\  {\bf 327}, 347 (2000)
 [Phys.\ Rept.\  {\bf 327}, 427 (2000)]
 [arXiv:hep-ph/9904229].

\bibitem{Vachaspati:1992fi}
 T.~Vachaspati,
 Phys.\ Rev.\ Lett.\  {\bf 68}, 1977 (1992)
 [Erratum-ibid.\  {\bf 69}, 216 (1992)].

\bibitem{James:1992zp}
 M.~James, L.~Perivolaropoulos and T.~Vachaspati,
 Phys.\ Rev.\ D {\bf 46}, 5232 (1992).


\bibitem{Forgacs:2005sf}
  P.~Forgacs, S.~Reuillon and M.~S.~Volkov,
  Phys.\ Rev.\ Lett.\  {\bf 96}, 041601 (2006)
  [arXiv:hep-th/0507246].

\bibitem{Forgacs:2006pm}
  P.~Forgacs, S.~Reuillon and M.~S.~Volkov,
  arXiv:hep-th/0602175.

\bibitem{Shi06}
M. Shifman and A. Yung, hep-th/0603134.

\bibitem{Blanco-Pillado:2005xx}
 J.~J.~Blanco-Pillado, G.~Dvali and M.~Redi,
 arXiv:hep-th/0505172.


\bibitem{Achucarro:1998ux}
 A.~Achucarro, J.~Borrill and A.~R.~Liddle,
 Phys.\ Rev.\ Lett.\  {\bf 82}, 3742 (1999)
 [arXiv:hep-ph/9802306].

\bibitem{Ach06}
A. Achucarro, P. Salmi and J. Urrestilla, astro-ph/0512487.

\bibitem{Benson:1993at}
 K.~Benson and M.~Bucher,
 Nucl.\ Phys.\ B {\bf 406}, 355 (1993)
 [arXiv:hep-ph/9304214].

\bibitem{Urrestilla:2004eh}
 J.~Urrestilla, A.~Achucarro and A.~C.~Davis,
 Phys.\ Rev.\ Lett.\  {\bf 92}, 251302 (2004)
 [arXiv:hep-th/0402032].

\bibitem{Dasgupta:2004dw}
 K.~Dasgupta, J.~P.~Hsu, R.~Kallosh, A.~Linde and M.~Zagermann,
 JHEP {\bf 0408}, 030 (2004)
 [arXiv:hep-th/0405247].

\bibitem{Mat88}
R.~A.~Matzner,
Computers in Physics, Sep/Oct, 51 (1988).

\bibitem{Moriarty:1988qs}
 K.~J.~M.~Moriarty, E.~Myers and C.~Rebbi,
{\it Proc. of Yale Workshop: Cosmic Strings: The Current Status,
eds. F.~S.~Acetta and L.~M.~Krauss,
New Haven, CT, May 6-7, 1988}

\bibitem{Laguna:1989hn}
 P.~Laguna and R.~A.~Matzner,
 Phys.\ Rev.\ Lett.\  {\bf 62}, 1948 (1989).

\bibitem{Shellard:1988ki}
 E.~P.~S.~Shellard,
{\it Proc. of Yale Workshop: Cosmic Strings: The Current Status,
eds. F.~S.~Acetta and L.~M.~Krauss,
New Haven, CT, May 6-7, 1988}

\bibitem{Bettencourt:1996qe}
 L.~M.~A.~Bettencourt, P.~Laguna and R.~A.~Matzner,
 Phys.\ Rev.\ Lett.\  {\bf 78}, 2066 (1997)
 [arXiv:hep-ph/9612350].

\bibitem{Vachaspati:1989xg}
 T.~Vachaspati,
 Phys.\ Rev.\ D {\bf 39}, 1768 (1989).

\bibitem{Nielsen:1973cs}
 H.~B.~Nielsen and P.~Olesen,
 Nucl.\ Phys.\ B {\bf 61}, 45 (1973).


\bibitem{Jackson:2004zg}
 M.~G.~Jackson, N.~T.~Jones and J.~Polchinski,
 arXiv:hep-th/0405229.

\bibitem{Hashimoto:2005hi}
 K.~Hashimoto and D.~Tong,
 arXiv:hep-th/0506022.

\bibitem{Vachaspati:1992pi}
 T.~Vachaspati and M.~Barriola,
 Phys.\ Rev.\ Lett.\  {\bf 69}, 1867 (1992).

\bibitem{Barriola:1993fy}
 M.~Barriola, T.~Vachaspati and M.~Bucher,
 Phys.\ Rev.\ D {\bf 50}, 2819 (1994)
 [arXiv:hep-th/9306120].


\bibitem{Gibbons:1992gt}
 G.~W.~Gibbons, M.~E.~Ortiz, F.~Ruiz Ruiz and T.~M.~Samols,
 Nucl.\ Phys.\ B {\bf 385}, 127 (1992)
 [arXiv:hep-th/9203023].

\bibitem{Achucarro:1992hs}
 A.~Achucarro, K.~Kuijken, L.~Perivolaropoulos and T.~Vachaspati,
 Nucl.\ Phys.\ B {\bf 388}, 435 (1992).

\bibitem{Abrikosov}
 A. A. Abrikosov,
 Zh.\ Eks.\ Th.\ F.\  {\bf 32}, 1442 (1957); JETP {\bf 5} 1174 (1957).


\bibitem{Ruback}
 P. J. Ruback, ``Vortex String Motion in the Abelian Higgs Model",
{\it Nucl. Phys.} {\bf 296} 669 (1988).

\end{thebibliography}
\end{document}